\documentclass[manuscript=letter,layout=traditional]{achemso}

\usepackage[version=3]{mhchem}
\usepackage{graphicx}
\usepackage{float}
\usepackage[english]{babel}
\usepackage{epstopdf}
\usepackage{amssymb}
\usepackage{pifont}
\usepackage{color}

\author{Csaba F\'{a}bri}
\email{ficsaba@caesar.elte.hu}
\affiliation{Laboratory of Molecular Structure and Dynamics, Institute of Chemistry, E\"otv\"os Lor\'and University, P\'azm\'any P\'eter s\'et\'any 1/A, H-1117 Budapest, Hungary}
\alsoaffiliation{MTA-ELTE Complex Chemical Systems Research Group, P.O. Box 32, H-1518 Budapest 112, Hungary}

\author{G\'abor J. Hal\'asz}
\affiliation{Department of Information Technology, University of Debrecen, P.O. Box 400, H-4002 Debrecen, Hungary}

\author{Lorenz S. Cederbaum}
\affiliation{Theoretische Chemie, Physikalisch-Chemisches Institut, Universit\"at Heidelberg, Im Neuenheimer Feld 229, 69120 Heidelberg, Germany}

\author{\'Agnes Vib\'ok}
\email{vibok@phys.unideb.hu}
\affiliation{Department of Theoretical Physics, University of Debrecen, PO Box 400, H-4002 Debrecen, Hungary}
\alsoaffiliation{ELI-ALPS, ELI-HU Non-Profit Ltd, H-6720 Szeged, Dugonics t\'er 13, Hungary}

\title{Born--Oppenheimer approximation in optical cavities: from success to breakdown}

\begin{document}

\begin{abstract}
The coupling of a molecule and a cavity induces nonadiabaticity in the molecule which makes the description of its dynamics complicated.
For polyatomic molecules, reduced-dimensional models and the use of the Born-Oppenheimer approximation (BOA) may remedy the situation.
It is demonstrated that contrary to expectation, BOA may even fail in a one-dimensional model and is generally expected to fail
in two- or more-dimensional models due to the appearance of conical intersections induced by the cavity.
\end{abstract}

\section{Introduction}

The dynamics initiated in a molecule by the absorption of a photon is usually
treated within the framework of the Born-Oppenheimer (BO) or adiabatic approximation
\cite{Born1} where the fast-moving electrons are separated
from the slow nuclear degrees of freedom (dofs). In this picture the nuclei move
on a single potential energy surface (PES) created by the fast-moving
electrons. Although several chemical processes can be rationalised by
considering a single BO PES, there are indeed a number of situations where the BO approximation (BOA)
breaks down. These are called nonadiabatic processes which involve nuclear
dynamics proceeding on at least two coupled PESs, leading to the formation of
so-called conical intersections (CIs).\cite{Lenz1,Yarkony,Baer1,Lenz2,Domcke1,Baer2}
Nonadiabatic phenomena are ubiquitous in photochemistry, photophysics,
particularly in molecular fragmentation, proton transfer, isomerization
or radiationless deactivation processes of excited states as the CI
can provide a very efficient channel for ultrafast interstate crossing on the femtosecond
timescale.\cite{Kim,Worner,You,Musser,Conta,Rebeca,10Martinez,16XiMaZh,16DeWeZh,17WoKaKi,18CuMa,18BeKoRo,17RyJoIz,19XiMaGu}

Conical intersections and avoided crossings (ACs) can be created both
by classical and quantum light, as well. To form a CI, the molecule must have
at least two independent nuclear dofs. In diatomics having only one nuclear dof,
natural CIs can never be formed, only ACs can arise. If the system interacts with light,
either light-induced avoided crossings (LIACs) or
light-induced conical intersections (LICIs)\cite{Lenz3,Gabi6,17CsHaCe} can
emerge. LICIs can be created even in diatomics where the 
second dof (either rotation or translation), needed to form a LICI,
comes into play due to the light-matter interaction.
Moreover, LICIs are ubiquitous and become multidimensional in the nuclear
coordinate space in polyatomic molecules due to the presence of several
vibrational dofs.\cite{Lenz5,20FaLaHa}

Recently, efforts have been made to study light-induced nonadiabatic phenomena in optical or microwave
cavities.\cite{15GaGaFe,16KoBeMu,17LuFeTo,17FlRuAp,17FlApRu,18DuMaRi,18RiMaDu,18FeGaGa,18Vendrell,Gabi10,19CsKoHa,19PeJuYu,19MaHu,19LiHe,19UlGoVe,20GuMu,20Herrera}
It has been successfully demonstrated that describing the photon-matter
interaction with the tools of cavity quantum electrodynamics (cQED)\cite{12HuScGe,16Ebbesen,16ChDeBe,16ZhChWa}
can provide an alternative way to study the quantum control of molecules with light.
In this framework nonadiabatic dynamics arises due to the strong coupling between the molecular dofs
and the photonic mode of the radiation field which can alter
the molecular energy levels by controlling the dynamics of basic photophysical
and photochemical processes. The molecular vibrational modes which
are strongly coupled to the electronic and photonic dofs
are taken into account resulting in a new set of ``cavity-induced''
or ``polariton'' surfaces in the molecular Hamiltonian.
These polariton surfaces are expected to form LIACs or LICIs.

Numerous works deal with quantum-light-induced nonadiabatic effects
within a single molecule. In most of the studies diatomic
or polyatomic organic molecules are treated as reduced-dimensional
two-level systems by taking into account only one vibrational and
photonic dof.\cite{16KoBeMu,15GaGaFe,17LuFeTo,18FeGaGa,18Vendrell,19UlGoVe}
As is already clear from classical light, quantum LICI situations can also only occur if,
in addition to the only vibrational dof, the rotational angle between the molecular
axis and the polarization vector of the electric field in the cavity is also accounted for
(in case of diatomics)\cite{Gabi10,19CsKoHa} or at least two vibrational
dofs are considered in the description.\cite{20FaLaHa,20GuMu}
Furthermore, quantum light-induced collective nonadiabatic phenomena (collective
LICI) can also emerge when many molecules are involved in strong coupling to 
the cavity mode.\cite{18FeGaGa,18Vendrell,19UlGoVe}

Our current aim is to study pure quantum light-induced nonadiabatic
phenomena in a single polyatomic molecule placed in an optical nanocavity
with methods ranging from a full-dimensional and accurate quantum-dynamical description to
a simple one-dimensional treatment. In order to eliminate any possible interference
between inherent and cavity-induced nonadiabatic phenomena, we
consider situations where a clear separation and identification of
these can be made, enabling us to reveal effects solely caused by the quantum LICI. 

Our showcase example is the four-atomic H$_{2}$CO (formaldehyde) molecule
which has been investigated very recently for nonadiabatic phenomena induced by classical light.\cite{20FaLaHa}
This molecule does not exhibit any inherent nonadiabatic effects in the studied region
of the nuclear configuration space (see further explanation in the Supporting Information).
Therefore, nonadiabatic effects appearing in the absorption spectrum of H$_{2}$CO
coupled to a single cavity mode can be attributed solely to the quantum LICI.
First, by applying accurate full-dimensional  computations, the field-free absorption spectrum
of H$_{2}$CO is investigated and compared to results obtained with one- and two-dimensional quantum-dynamical models.
Next, we study the absorption spectrum of H$_{2}$CO coupled to a single cavity mode
and investigate how cavity-induced nonadiabatic phenomena can be understood
if simplified one- and two-dimensional descriptions are used.
Most importantly, we examine whether the BOA is capable of yielding qualitatively (or even quantitatively)
correct absorption spectra for the different quantum-dynamical models applied in this study.

The BOA, to be accurately defined later in the context of polaritonic surfaces,
breaks down for both the two- and six-dimensional quantum-dynamical models used in this work.
The failure of the BOA is attributed to the emergence of the LICI between polaritonic surfaces.
However, if only one vibrational dof is taken into account, no LICI can be formed and it seems
plausible that the BOA can provide correct absorption spectra.
In ref. \citenum{15GaGaFe} it has been found that the BOA is valid for larger organic molecules described
in one dimension provided that the coupling between the molecule and the cavity mode is sufficiently strong.
The most striking outcome of the present work is that the BOA can fail even for a one-dimensional description
of H$_2$CO, irrespective of the coupling strength.
Consequently, BOA is expected to fail in the presence of a LICI and care should be taken
when a molecule coupled to a cavity mode is described using the BOA with only one vibrational dof.

\section{Results and discussion}

\subsection{Hamiltonian of a single molecule coupled to a cavity mode and six-dimensional results}

Let us start with the Hamiltonian of a single molecule coupled to a cavity mode,\cite{04CoDuGr}
\begin{equation}
	\hat{H} = \hat{H}_0 + \hbar \omega_\textrm{c} \hat{a}^\dag \hat{a} - g \hat{\vec{\mu}} \vec{e} (\hat{a}^\dag + \hat{a})
\end{equation}
where $\hat{H}_0$ denotes the Hamiltonian of the isolated molecule, $\hat{\vec{\mu}}$ is the electric dipole moment operator,
$\vec{e}$ is the polarization vector, $\hat{a}^\dag$ and $\hat{a}$ are the creation and annihilation operators of the cavity mode,
$\omega_\textrm{c}$ is the angular frequency of the cavity mode and $g$ is the coupling strength parameter.
If two electronic states of the molecule are considered, the Hamiltonian takes the form
\begin{equation}
    \resizebox{0.9\textwidth}{!}{$\hat{H} = 
         \begin{bmatrix}
            \hat{T} + V_\textrm{X} & 0 & 0 & W_1 & 0 & 0 & \dots \\
            0 & \hat{T} + V_\textrm{A} & W_1 & 0 & 0 & 0 & \dots \\
            0 & W_1 & \hat{T} + V_\textrm{X} + \hbar\omega_\textrm{c} & 0 & 0 & W_2 & \dots \\
            W_1 & 0 & 0 &\hat{T} + V_\textrm{A} + \hbar\omega_\textrm{c} & W_2 & 0 & \dots \\
            0 & 0 & 0 & W_2 &\hat{T} + V_\textrm{X} + 2\hbar\omega_\textrm{c} & 0 & \dots \\
            0 & 0 & W_2 & 0 & 0 &\hat{T} + V_\textrm{A} + 2\hbar\omega_\textrm{c} & \dots \\
            \vdots & \vdots & \vdots & \vdots & \vdots & \vdots & \ddots 
        \end{bmatrix}$}
    \label{eq:cavity_H}
\end{equation}
where $\hat{T}$ is the kinetic energy operator, $V_\textrm{X}$ and $V_\textrm{A}$ denote
the ground-state and excited-state PESs and  $W_n = -g  \sqrt{n} \vec{\mu} \vec{e}$,
where $\vec{\mu}$ is the transition dipole moment (TDM) vector.
In what follows, the notation $\mu = \vec{\mu} \vec{e}$ will be used.

The eigenstates of the Hamiltonian of Eq. \eqref{eq:cavity_H} 
\begin{equation}
	| \Phi_k \rangle = \sum_{\alpha=\textrm{X},\textrm{A}} \sum_i \sum_n c^{(k)}_{\alpha i n} | \alpha i \rangle |n\rangle
\label{eq:dressed_state}
\end{equation}
can be obtained as the linear combination of the products of field-free molecular vibronic eigenstates (denoted by 
$| \textrm{X}i \rangle$ and $| \textrm{A}i \rangle$) and Fock states $| n \rangle$ ($n=0,1,2,\dots$) of the cavity mode.
The intensities of electric dipole transitions between the eigenstates $| \Phi_k \rangle$ and $| \Phi_l \rangle$ are expressed as
\begin{equation}
	I_{kl} \propto \omega_{kl} \sum_{\alpha=x,y,z} |\langle \Phi_k | \hat{\mu}_\alpha | \Phi_l \rangle|^2
\end{equation}
where $\omega_{kl}$ is the angular frequency of the transition and
$\hat{\mu}_\alpha$ denotes the components of the electric dipole moment operator.

\begin{figure}[hbt!]
 \centering
   \includegraphics[scale=0.35]{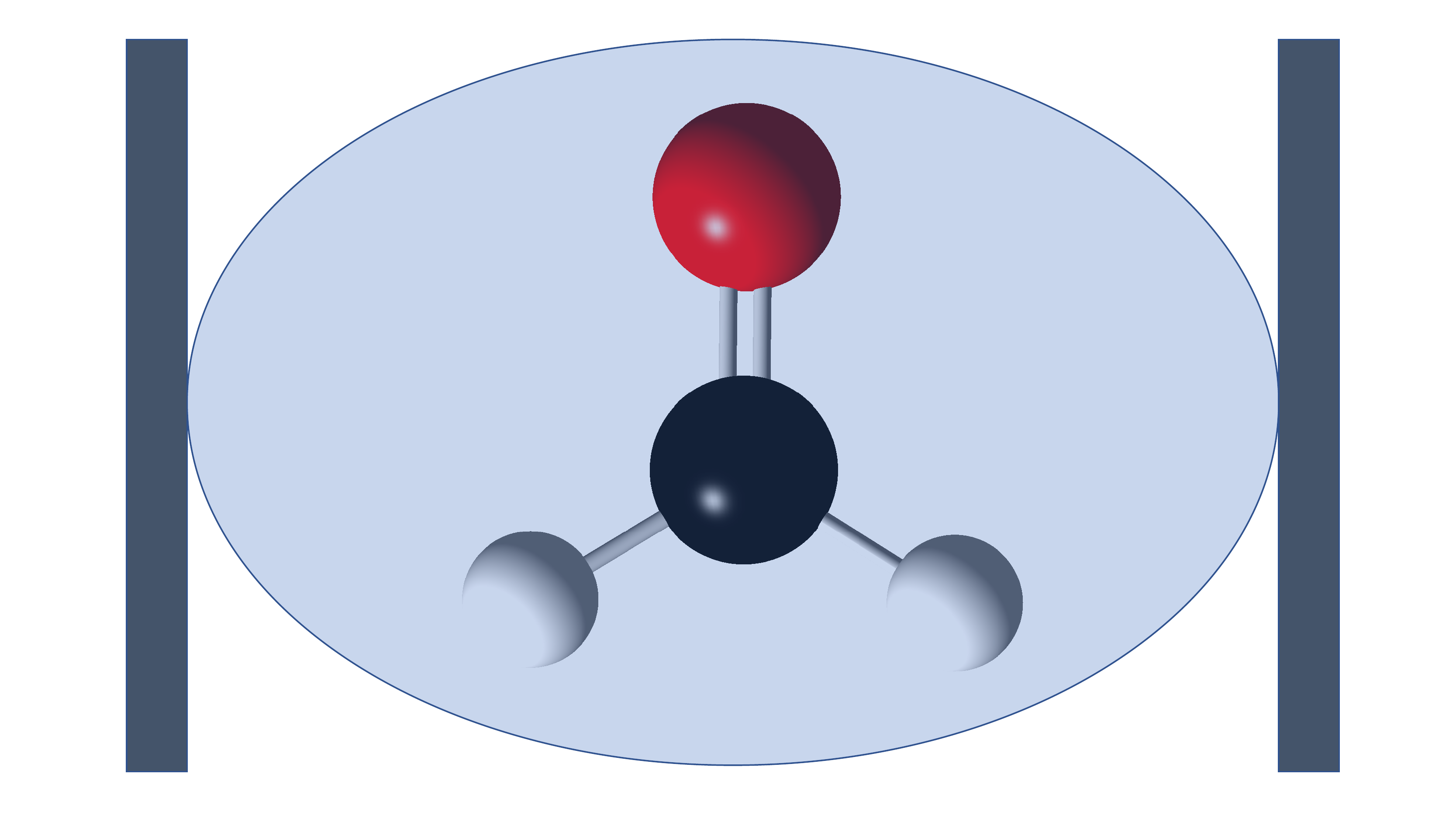}
   \includegraphics[scale=0.85]{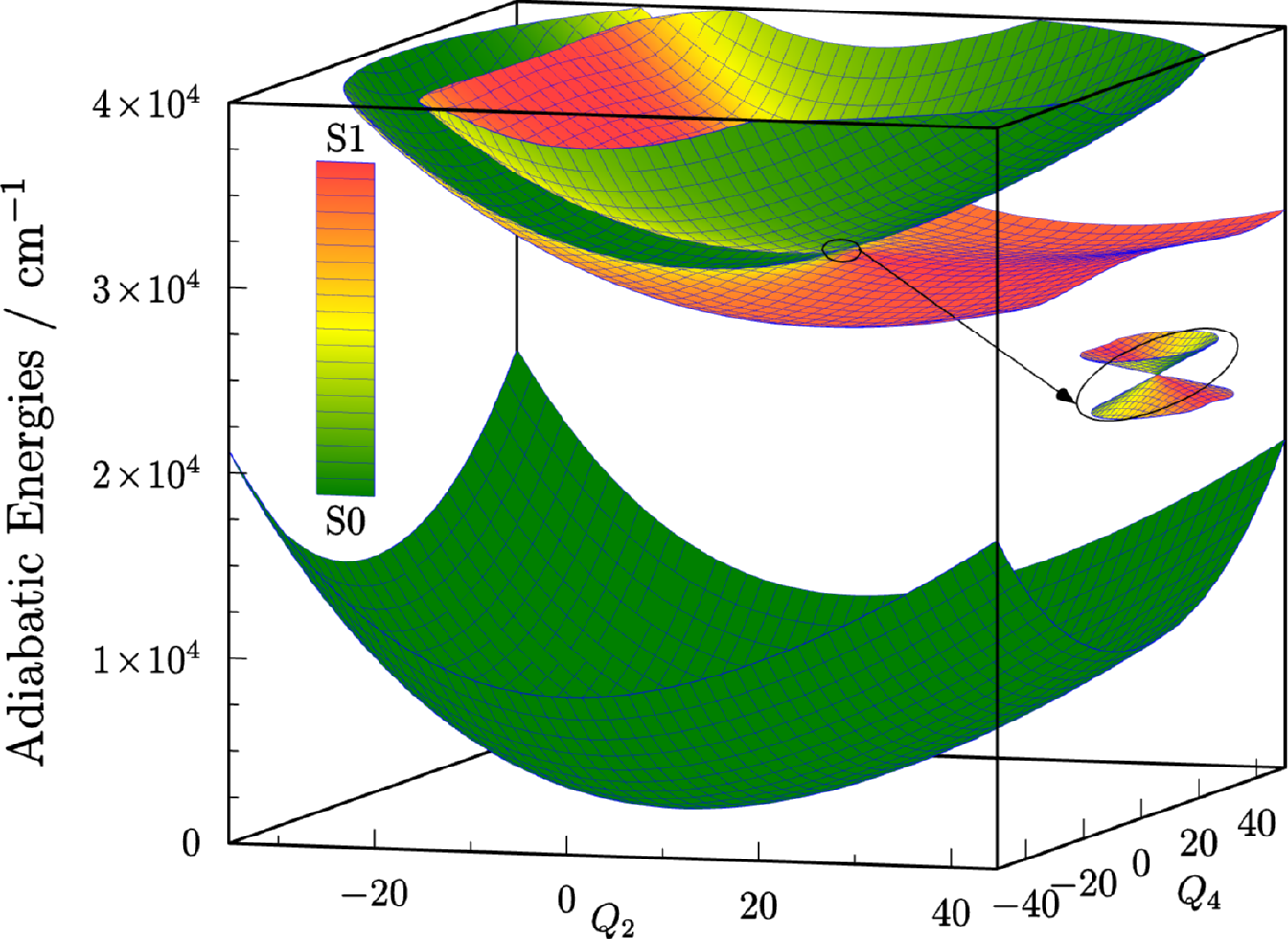}
   \caption{Structure of the H$_2$CO (formaldehyde) molecule (upper panel) and the three lowest
   		polaritonic (adiabatic) surfaces of H$_2$CO coupled to a single cavity mode (lower panel).
		$Q_2$ and $Q_4$ are the normal coordinates of the $\nu_2$ (C=O stretch) and
		$\nu_4$ (out-of-plane bend) vibrational modes. The cavity wavenumber and coupling strength
		are chosen as $\omega_\textrm{c} = 29000 ~ \textrm{cm}^{-1}$ and $g = 5.97 \cdot 10^{-2} ~ \textrm{au}$, respectively.
		The light-induced conical intersection between the second and third
		polaritonic surfaces is shown in the inset on the right-hand side of the lower panel.
		The character of the polaritonic surfaces is indicated by different colours (see the legend on the left).}
   \label{fig:molecule_cavity}
\end{figure}

Following the footsteps of our previous study focused on light-induced nonadiabaticity in polyatomic molecules,\cite{20FaLaHa}
we choose the four-atomic H$_2$CO molecule as the target of the present work.
The two singlet electronic states $\textrm{S}_0 ~ (\tilde{\textrm{X}} ~ ^1\textrm{A}_1)$ and
$\textrm{S}_1 ~ (\tilde{\textrm{A}} ~ ^1\textrm{A}_2)$ of H$_{2}$CO are taken into account and
the corresponding six-dimensional $V_{\textrm{X}}$ and $V_{\textrm{A}}$ PESs are taken from
refs. \citenum{Bowman2} and \citenum{Bowman1}, respectively.
The structure of H$_2$CO coupled to a single cavity mode and the three lowest polaritonic (adiabatic)
PESs that emerge due to the light-matter coupling are depicted in Fig. \ref{fig:molecule_cavity}.
The field-free vibrational eigenstates
of H$_{2}$CO were computed with the GENIUSH program package\cite{09MaCzCs,11FaMaCs,12CsFaSz}
for both electronic states treating all six vibrational dofs in a numerically exact way.
The rotational dofs are omitted from the present computational protocol and the molecule
is imagined to be fixed with respect to the external electric field.
Further information on the structure and normal modes of H$_2$CO as well as technical details
of the computations are provided in the Supporting Information.

\begin{figure}[hbt!]
 \centering
   \includegraphics[scale=0.8]{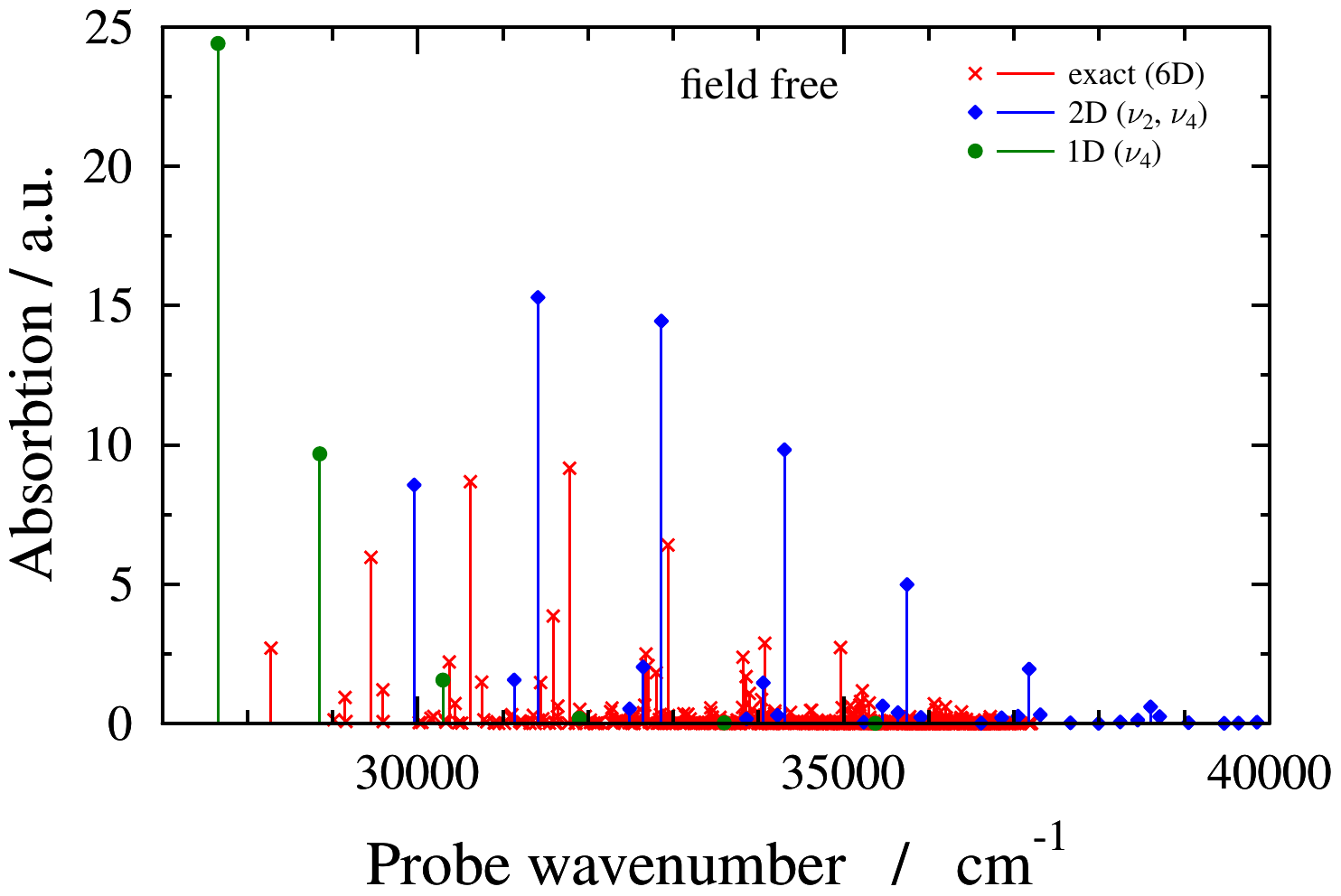}
   \caption{Absorption spectra of field-free (i.e., no cavity) formaldehyde obtained with different quantum-dynamical models
            (6D, 2D($\nu_2$,$\nu_4$) and 1D($\nu_4$) models, as indicated in the figure).}
   \label{fig:spectra_ff}
\end{figure}

Having described the working formulae and the molecular system, we move on to the discussion of
the absorption spectrum of the field-free H$_2$CO molecule and first present results with a numerically-exact six-dimensional (6D) quantum-dynamical treatment.
We consider the high-energy part of the spectrum consisting of spectral lines corresponding to transitions from $|\textrm{X}0\rangle$
(vibrational ground state of the electronic state X) to $|\textrm{A}i\rangle$ (vibrational states of the electronic state A).
Fig. \ref{fig:spectra_ff} presents the 6D field-free spectrum of H$_2$CO, showing favourable agreement
with results reported in ref. \citenum{18BoPeEi}. The 6D field-free spectrum exhibits progressions of lines that are mainly
associated with the $\nu_4$ (out-of-plane bend) and $\nu_2$ (C=O stretch) vibrational modes. Therefore, any reduced-dimensional
quantum-dynamical model of H$_2$CO should incorporate at least the $\nu_4$ and $\nu_2$ vibrational modes.

Fig. \ref{fig:spectra_ff} also shows field-free spectra obtained with two reduced-dimensional models, treating the
$\nu_2$ and $\nu_4$ vibrational modes (2D($\nu_2,\nu_4$) model), or the $\nu_4$ vibrational mode (1D($\nu_4$) model).
Comparing the field-free 2D($\nu_2,\nu_4$) spectrum to its 6D counterpart reveals that although the 2D($\nu_2,\nu_4$) spectrum
lacks lines that are present in the 6D spectrum, the overall structures of the 2D($\nu_2,\nu_4$) and 6D field-free spectra are similar.
On the contrary, the 1D($\nu_4$) field-free spectrum does not resemble the 6D field-free spectrum at all.
Besides the 1D($\nu_4$) model, one could also think of testing the performance of the 1D($\nu_2$) model.
Since the reduced-dimensional models are defined by setting the inactive normal coordinates to zero and the $\nu_2$
vibrational mode is totally symmetric, the $C_{2v}$ symmetry of the equilibrium structure is preserved when displacements are
made along the $\nu_2$ vibrational mode (see also the Supporting Information). Due to symmetry, the TDM between the electronic states X and A
is zero at nuclear configurations of $C_{2v}$ symmetry, therefore, no transitions are allowed if the 1D($\nu_2$) model is used.
Based on the analysis of the field-free spectra presented in this subsection, the simplest model expected to
provide sensible results is the 2D($\nu_2$,$\nu_4$) model.

\subsection{One-dimensional results for the 1D($\nu_4$) model}
We define a one-dimensional quantum-dynamical model, called 1D($\nu_4$) model henceforth,
treating only the $\nu_4$ (out-of-plane) vibrational mode of H$_2$CO (see the Supporting Information for technical details).
The 1D($\nu_4$) model is utilized to assess the performance of a simplified one-dimensional description for H$_2$CO and
test the applicability of the BOA in one dimension, which requires further explanation.
Since the frequency of the cavity mode is chosen to be nearly resonant with the frequency of the 
$\textrm{X} \rightarrow \textrm{A}$ electronic excitation, the dof associated with the ``fast'' cavity mode is
grouped with the electrons and the ``slow'' nuclear dofs are separated from the electronic and cavity dofs.
This allows the construction of the lower ($|-\rangle$) and upper ($|+\rangle$) hybrid light-matter (polaritonic) states
that can be approximately described as
\begin{align}
	|-\rangle = a | \textrm{X} \rangle |1\rangle - b | \textrm{A} \rangle |0\rangle \nonumber \\
	|+\rangle = b | \textrm{X} \rangle |1\rangle + a | \textrm{A} \rangle |0\rangle
\end{align}
in the singly-excited subspace.
The corresponding polaritonic (adiabatic) PESs can be obtained as eigenvalues of the potential energy part
of the Hamiltonian of Eq. \eqref{eq:cavity_H} at each nuclear configuration.
Throughout this work, the BOA is defined by neglecting the nonadiabatic coupling (NAC) between
the lower and upper polaritonic PES ($V_-$ and $V_+$). Next, the absorption spectrum is computed using
the BOA and the BOA spectrum is compared to the exact absorption spectrum that includes all effects caused by the NAC.
Note that all spectra presented in this subsection have been calculated utilizing the 1D($\nu_4$) model.

The initial dressed state $| \Phi_\textrm{i} \rangle$ is always chosen as the lowest-energy dressed state.
As for the current range of $g$ the lowest polariton PES (see the first polariton PES in Fig. \ref{fig:molecule_cavity})  can be approximated as
$V_\textrm{X}$ (strong coupling regime), $| \Phi_\textrm{i} \rangle$ virtually equals $| \textrm{X}0 \rangle |0\rangle$, that is,
the product of $| \textrm{X}0 \rangle$ and the vacuum state of the cavity mode.
The final dressed states $| \Phi_\textrm{f} \rangle$ of the transitions lie in the singly-excited subspace
(molecule in ground electronic state X dressed with one photon or molecule in excited electronic state A dressed with zero photon).
Note that the second and third polaritonic PESs in Fig. \ref{fig:molecule_cavity} correspond to the singly-excited subspace.
We stress that in the numerical computations all coupling terms appearing in the Hamiltonian of
Eq. \ref{eq:cavity_H} are included explicitly.

\begin{figure}
 \centering
   \includegraphics[scale=0.7]{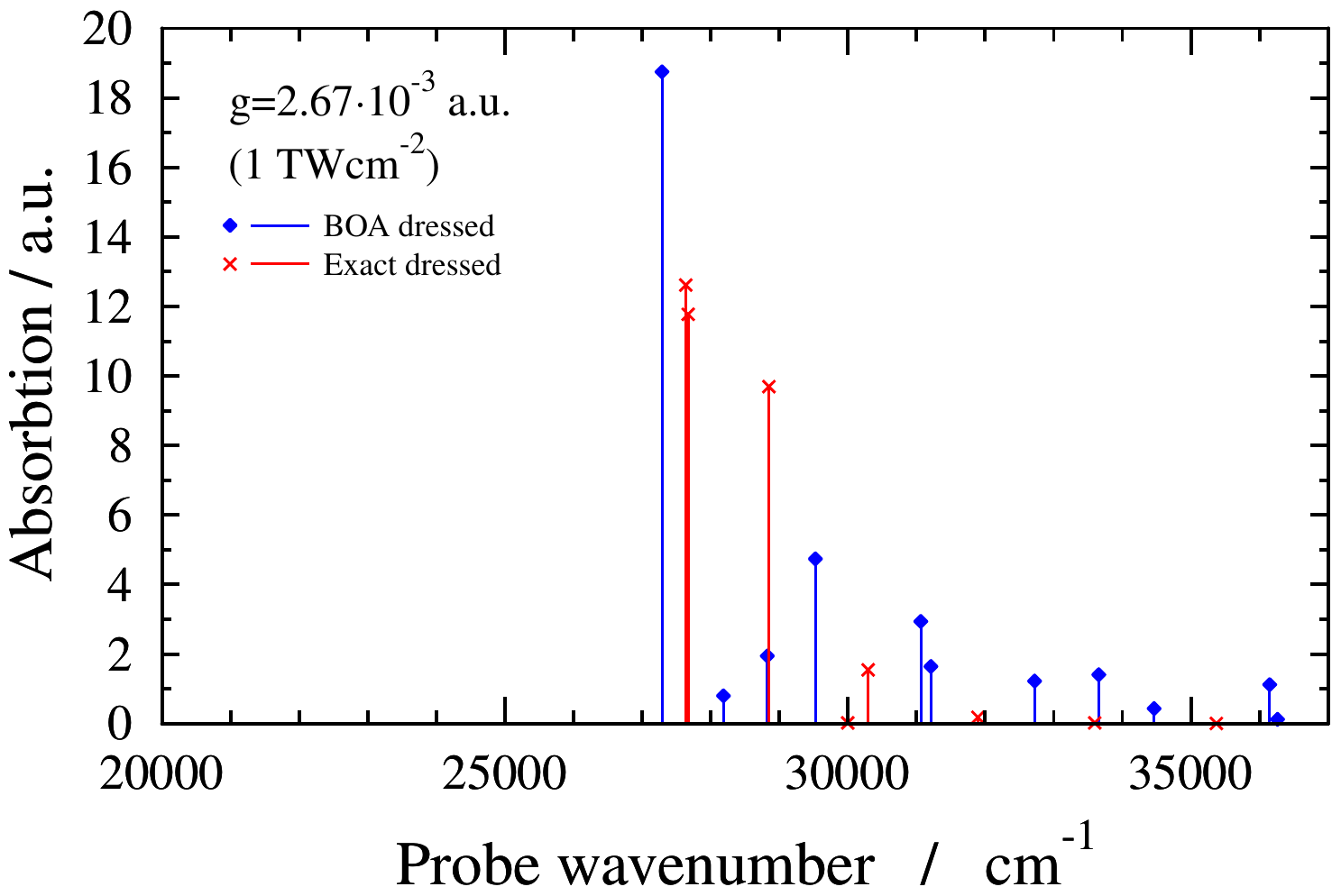}
   \includegraphics[scale=0.7]{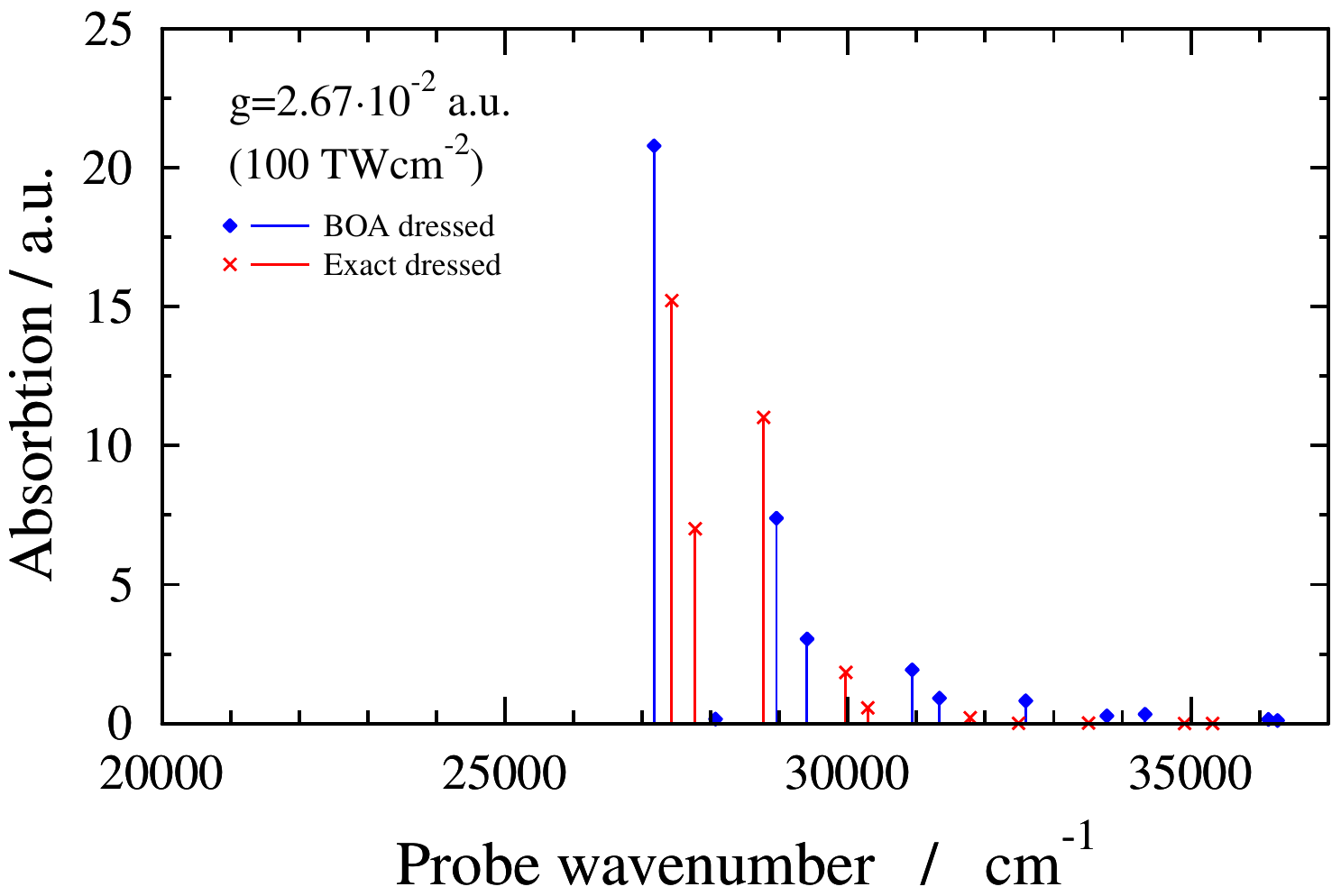}
   \includegraphics[scale=0.7]{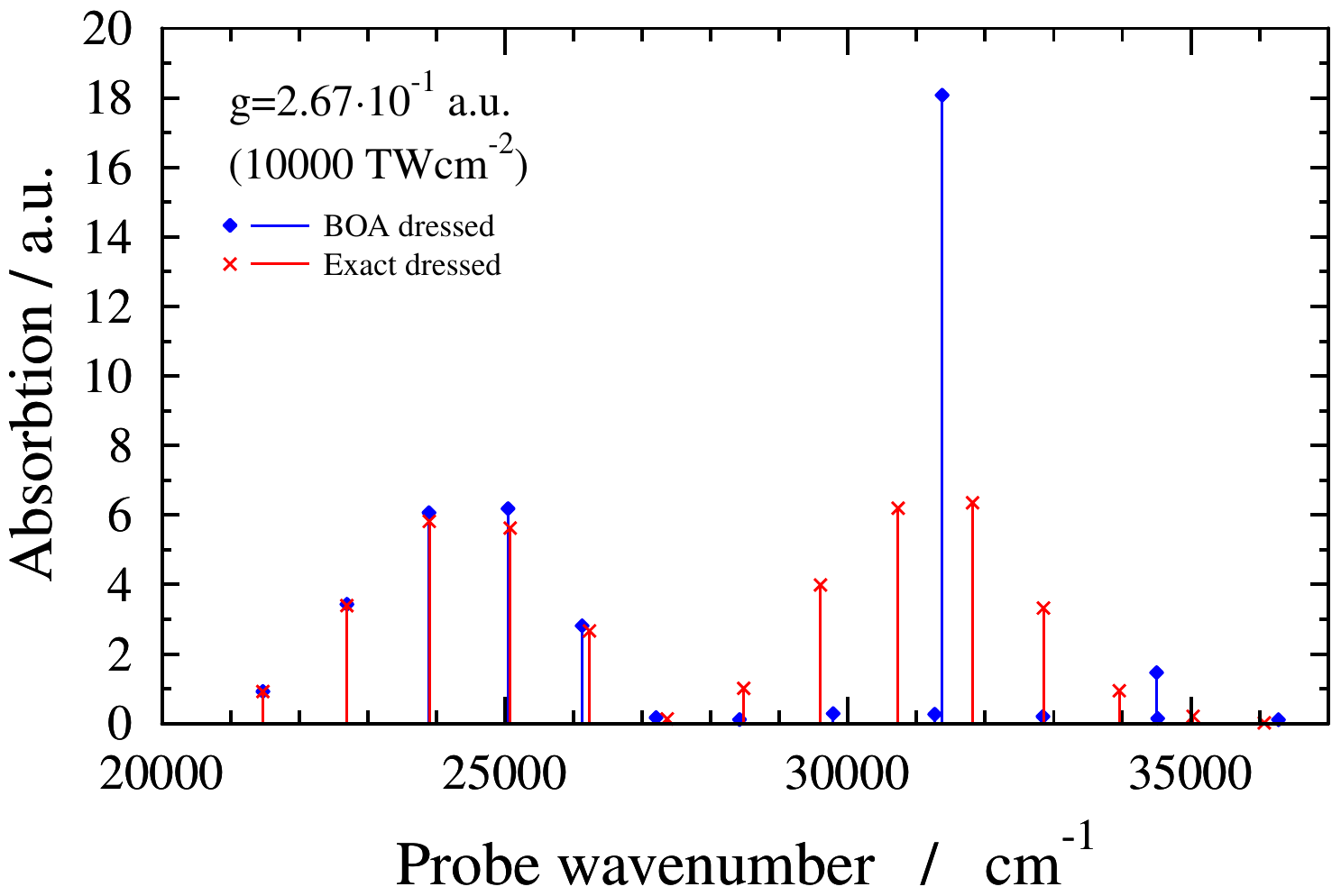}
   \caption{1D($\nu_4$)  dressed spectra (exact and BOA) with a cavity wavenumber of $\omega_\textrm{c} = 27653.3 ~ \textrm{cm}^{-1}$ for different coupling strength values.}
   \label{fig:spectra_1D}
\end{figure}

The exact dressed spectra displayed in Fig. \ref{fig:spectra_1D} ($\omega_\textrm{c} = 27653.3 ~ \textrm{cm}^{-1}$ and $\mathbf{e} = (0,1,0)$
with different values of $g$) show the emergence of new peaks besides splittings and shifts of peaks that are
present in the field-free spectrum. It is apparent in Fig. \ref{fig:spectra_1D}  that the dressed spectra calculated using the BOA
are rather different from their exact counterparts for all $g$ values considered.
The lower panel in Fig. \ref{fig:spectra_1D} shows the dressed spectrum with $g = 2.67 \cdot 10^{-1} ~ \textrm{au}$
(equivalent to a classical intensity of $I=10000 ~ \textrm{TW cm}^{-2}$), the highest (and likely experimentally not yet feasible) value of $g$ applied.
In contrast to the dressed spectra with lower $g$ values, the spectrum with $g = 2.67 \cdot 10^{-1} ~ \textrm{au}$ exhibits two clearly
separated groups of peaks. Moreover, it is conspicious that the BOA works for the lower group of peaks and fails completely
 for the upper group of peaks in this particular case.
 
\begin{figure}
 \centering
   \includegraphics[scale=0.525]{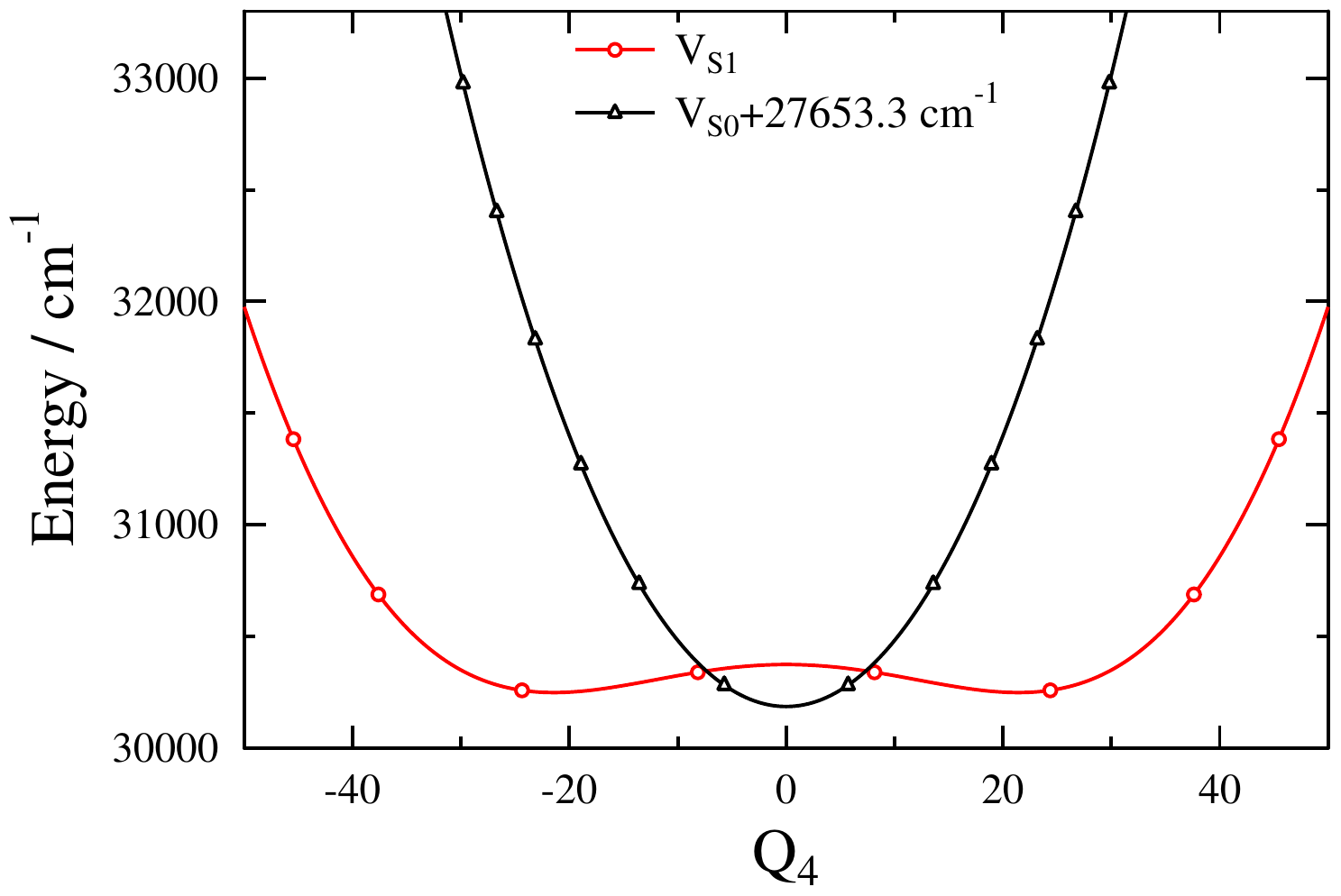}
   \includegraphics[scale=0.525]{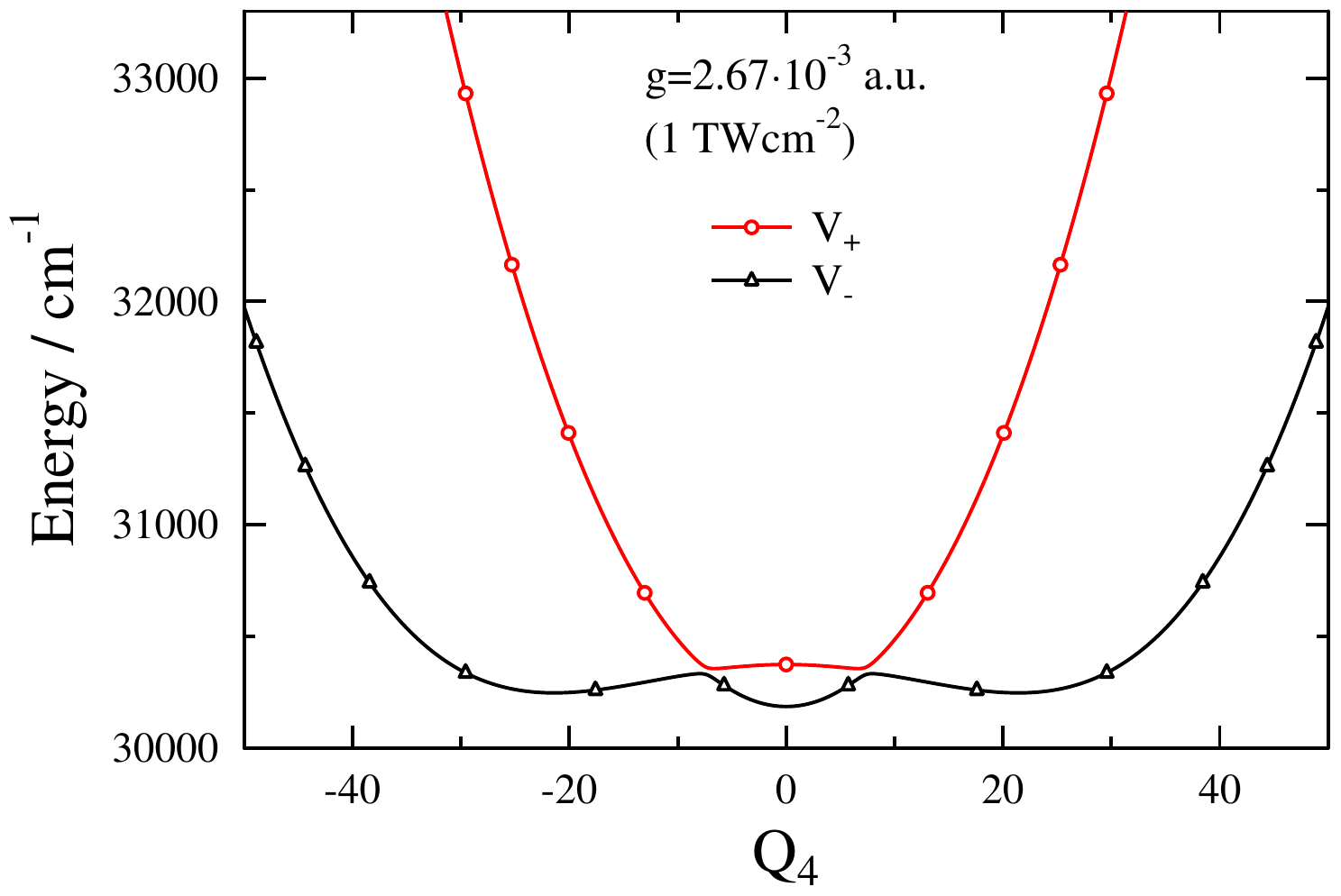}
   \includegraphics[scale=0.525]{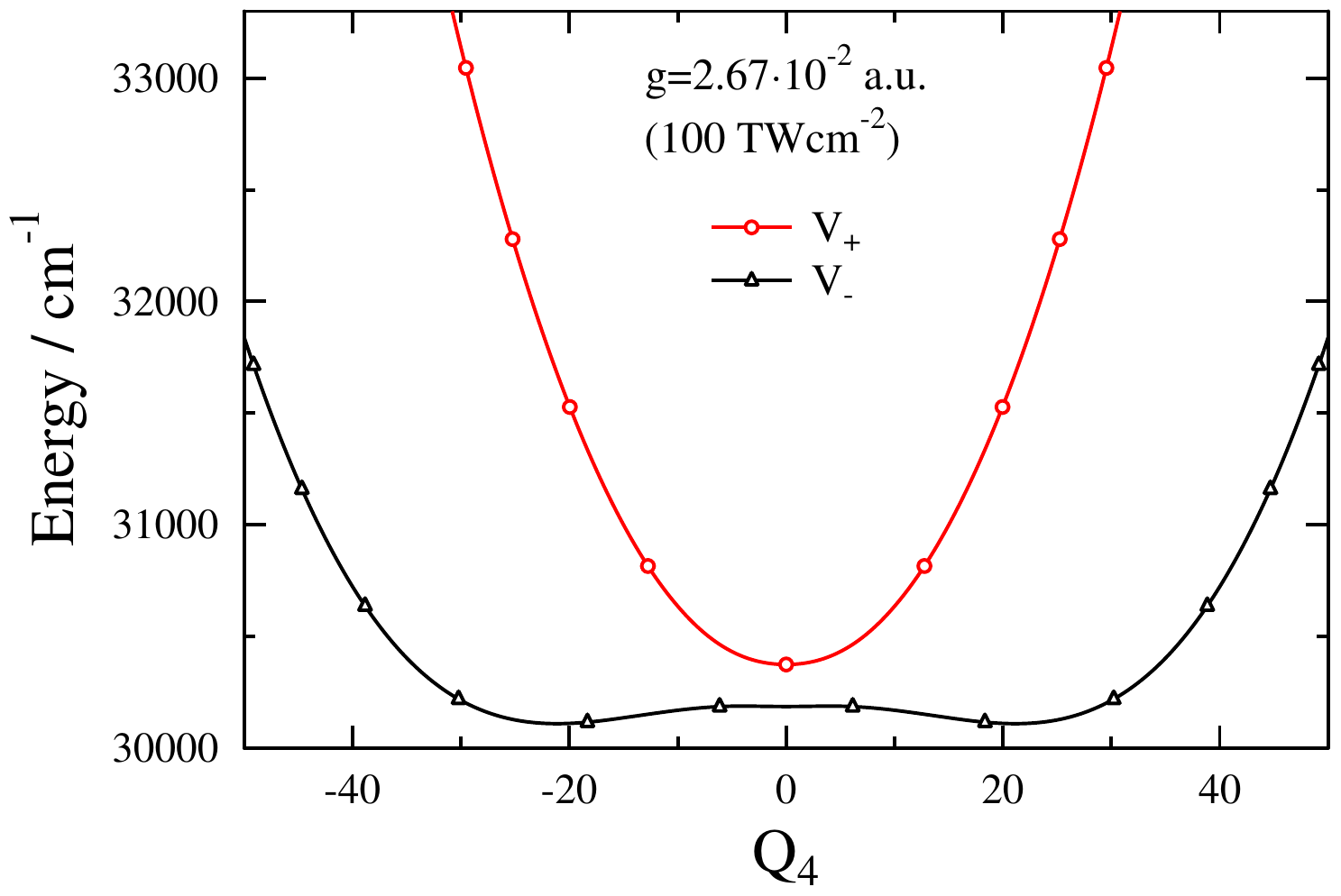}
   \includegraphics[scale=0.525]{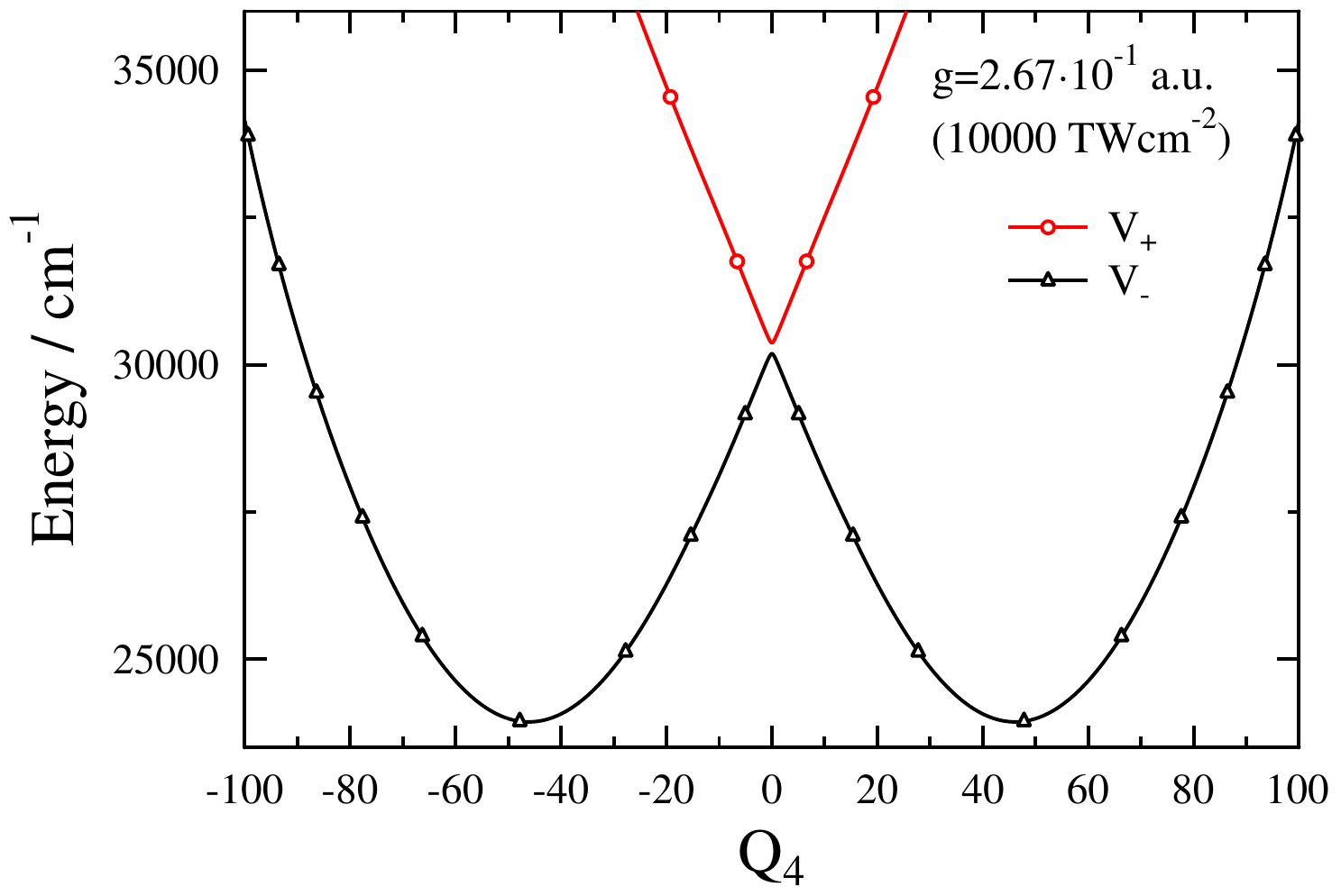}
   \caption{1D($\nu_4$) diabatic (upper left panel) and adiabatic potential curves ($V_+$ and $V_-$) with a
            cavity wavenumber of $\omega_\textrm{c} = 27653.3 ~ \textrm{cm}^{-1}$ for different coupling strength values.}
   \label{fig:pes_1D}
\end{figure}

\begin{figure}
 \centering
   \includegraphics[scale=0.8]{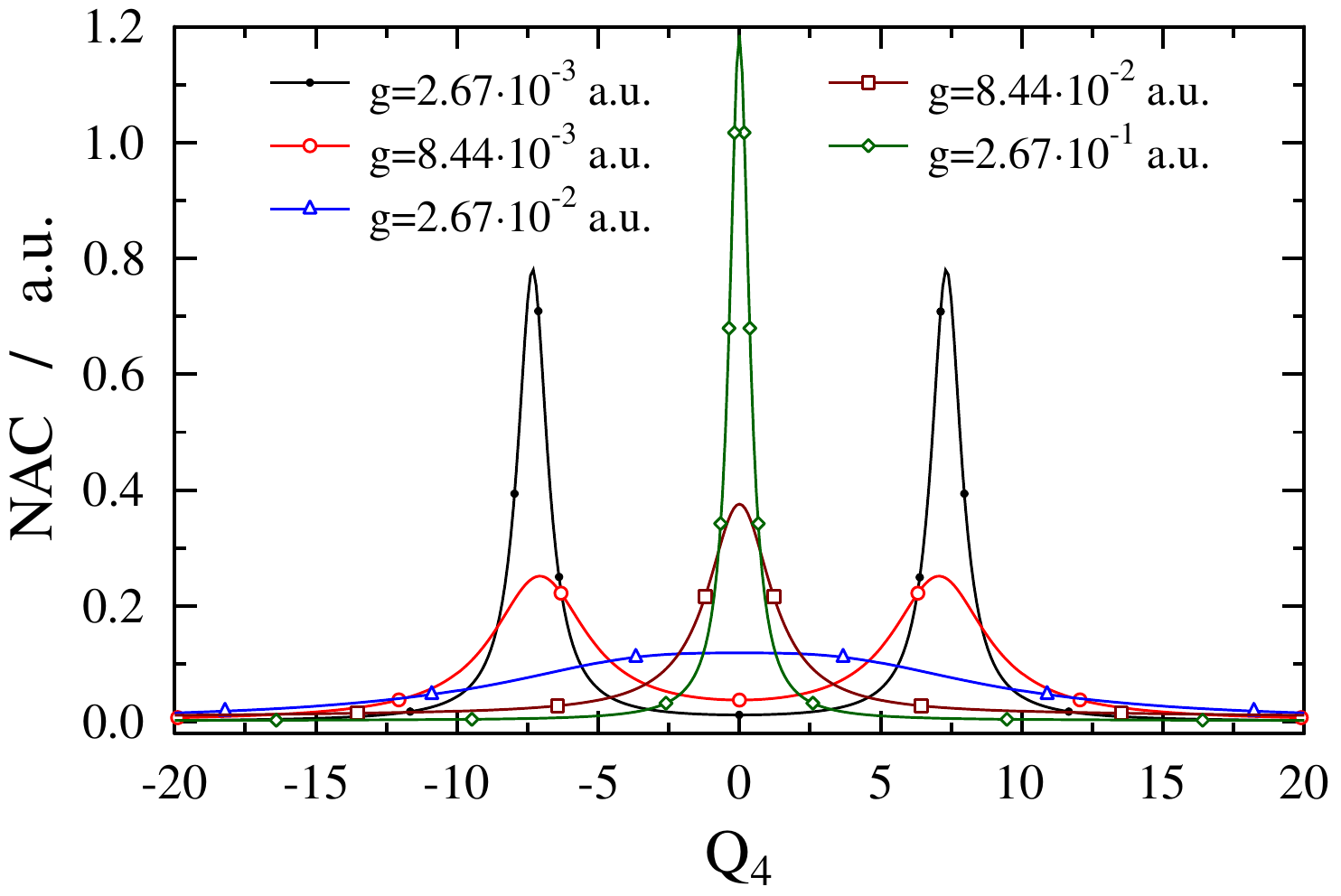}
   \caption{Nonadiabatic coupling (1D($\nu_4$) model) as a function of $Q_4$ for formaldehyde in a cavity  
   	 	of $\omega_\textrm{c} = 27653.3 ~ \textrm{cm}^{-1}$ with different coupling strength values $g$.
		With growing coupling strength, the bimodal structure of the nonadiabatic coupling turns into a
		single maximum following the structure of the avoided crossings of the adiabatic potentials shown in Fig. \ref{fig:pes_1D}.}
   \label{fig:nac_1D}
\end{figure}

The breakdown of the BOA even for the highest $g$ value might seem counterintuitive as only one vibrational dof is taken into account.
In order to rationalise this odd result we have investigated the lower and upper polaritonic (adiabatic) PESs and evaluated the NAC for several $g$ values.
Fig. \ref{fig:pes_1D} displays both the diabatic ($V_\textrm{X} + \hbar\omega_\textrm{c}$ and $V_\textrm{A}$) and adiabatic PESs ($V_-$ and $V_+$) as a function
of the $\nu_4$ normal coordinate ($Q_4$). The diabatic PESs cross at $Q_4 = \pm 7.35$ and $30344.8 ~ \textrm{cm}^{-1}$ for $\omega_\textrm{c} = 27653.3 ~ \textrm{cm}^{-1}$.
Note that $V_\textrm{A}$ shows an anharmonic double-well structure
and the TDM vanishes at $Q_4 = 0$ due to symmetry, which implies that the gap between $V_-$ and $V_+$ at $Q_4 = 0$
is determined solely by $\omega_\textrm{c}$ (and not by $g$).
One can also observe LIACs in Fig. \ref{fig:pes_1D} and the shapes of $V_-$ and $V_+$ change substantially as $g$ increases.
An interesting feature of $V_-$ is the emergence of a barrier centered at $Q_4 = 0$ for high $g$ values.
Fig. \ref{fig:nac_1D} shows the NAC as a function of $Q_4$ for several $g$ values. One can notice in Fig. \ref{fig:nac_1D} that
at lower $g$ values the NAC curve has a bimodal structure (with maxima located around the two crossing points of the diabatic PESs),
while at higher $g$ values the NAC exhibits a single maximum at $Q_4 = 0$.

The numerical NAC results can be readily interpreted by a simple one-dimensional model with two harmonic PESs,
\begin{align}
	V_\textrm{X}(x) &= \frac{1}{2} m \omega_\textrm{X}^2 x^2 \nonumber \\
	V_\textrm{A}(x) &= \frac{1}{2} m \omega_\textrm{A}^2 x^2 + \Delta,
\label{eq:modelPES}
\end{align}
coupled by a linear TDM,
\begin{equation}
	\mu(x) = \alpha x,
\end{equation}
which, similarly to the TDM of the 1D($\nu_4$) model, vanishes at $x=0$.
In Eq. \eqref{eq:modelPES} $m$ is the mass of the oscillator, $\omega_\textrm{X}$ and 
$ \omega_\textrm{A}$ refer to the harmonic frequencies of the ground and excited
electronic states and $\Delta$ denotes the excitation energy. 
Although the 1D($\nu_4$) $V_\textrm{A}$ PES has an anharmonic double-well structure, we believe
that the harmonic approximation for $V_\textrm{A}$ still yields a correct interpretation of the
$g$-dependence of the NAC.
As the $\textrm{NAC}(x)$ formula is rather involved (see the Supporting Information for details of the analytical derivations),
here the NAC is evaluated only at the two crossing points of the two diabatic PESs ($x_0$), that is,
\begin{equation}
    \textrm{NAC}(x_0) = \frac{m (\omega_\textrm{X}^2-\omega_\textrm{A}^2)}{4 g \alpha} \propto g^{-1},
   \label{eq:1Dnac_x0}
\end{equation}
and at $x=0$, where one gets
\begin{equation}
    \textrm{NAC}(0) = \frac{g \alpha}{\Delta-\hbar\omega_\textrm{c}} \propto g.
    \label{eq:1Dnac_0}
\end{equation}
While Eq. \eqref{eq:1Dnac_x0} shows that the NAC is inversely proportional to $g$ around $x_0$,
implying that the NAC becomes negligible for sufficiently large $g$ values in this region,
Eq. \eqref{eq:1Dnac_0} clearly indicates that the NAC is proportional to $g$ at $x = 0$.
This striking behaviour of the NAC provides an explanation for the shape of the NAC curves in Fig. \ref{fig:nac_1D} and
indicates that the BOA indeed breaks down even for large $g$ values in the 1D($\nu_4$) model.

Finally, the interpretation of the partial BOA breakdown for the highest value of $g$  ($2.67 \cdot 10^{-1} ~ \textrm{au}$) is in order.
The analysis of the dressed spectrum reveals that the peaks of the lower group in the BOA spectrum (see lower panel in Fig. \ref{fig:spectra_1D})
correspond to transitions to the lower-lying eigenstates of $V_-$. Since $V_-$ exhibits a high barrier in this case (see lower right panel in Fig. \ref{fig:pes_1D}),
the lower-lying adiabatic eigenstates of $V_-$ have negligible amplitude around $Q_4 = 0$, i.e., in the region where the NAC is not negligible
according to Fig. \ref{fig:nac_1D}. Therefore, the BOA provides satisfactory results for the lower group of peaks.
As to the higher group of peaks, the final dressed states of these BOA transitions are adiabatic eigenstates of $V_+$ which has a single minimum
at $Q_4 = 0$, therefore, the effect of the NAC can not be neglected and the BOA breaks down.

Interestingly, the higher group of states can be explained by the potential curve obtained by connecting smoothly the
left (right) part of $V_+$ with the right (left) part of $V_-$ in the lower right panel of Fig. \ref{fig:pes_1D}. 

\subsection{Two-dimensional results for the 2D($\nu_2,\nu_4$) model}
\begin{figure}
 \centering
   \includegraphics[scale=0.7]{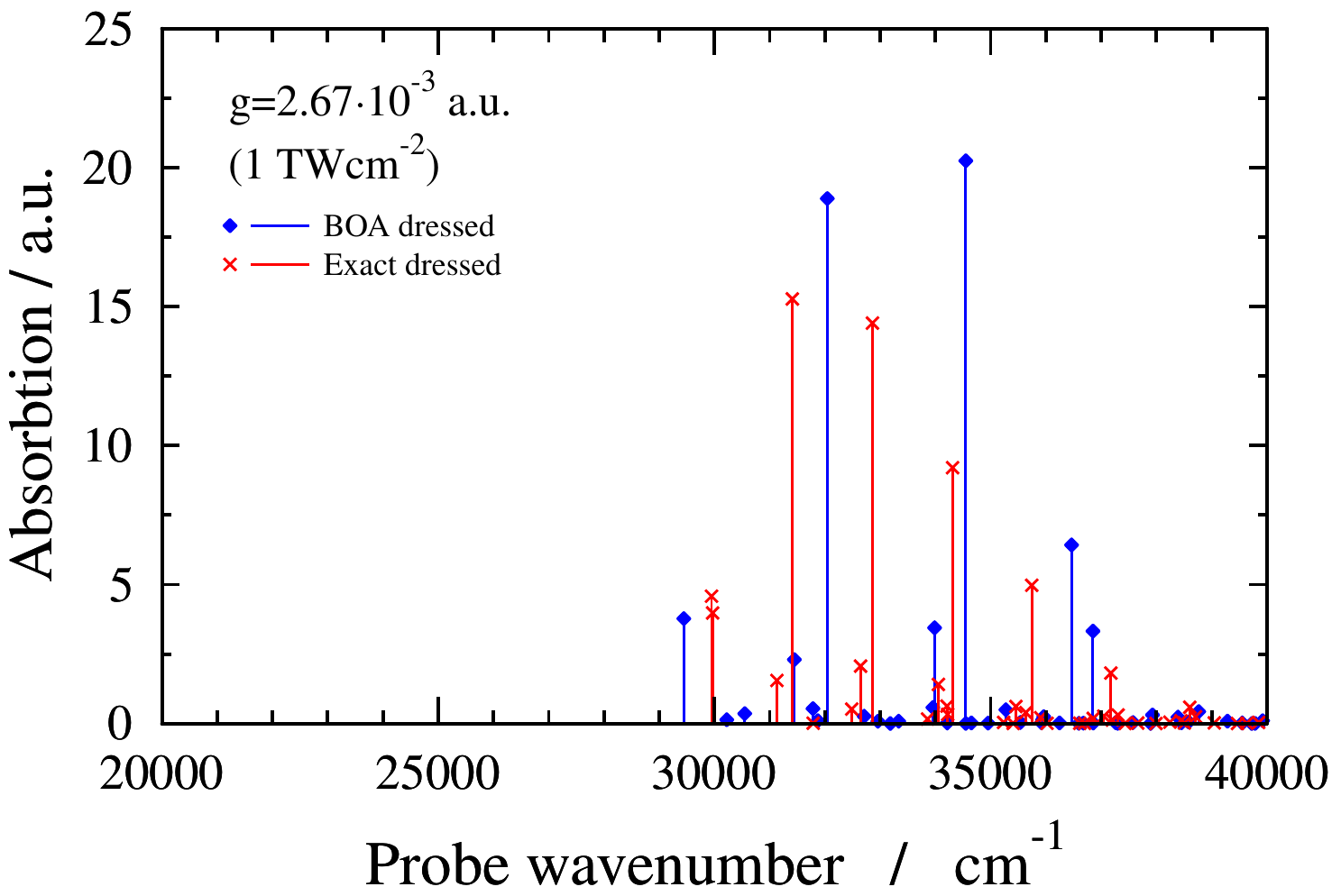}
   \includegraphics[scale=0.7]{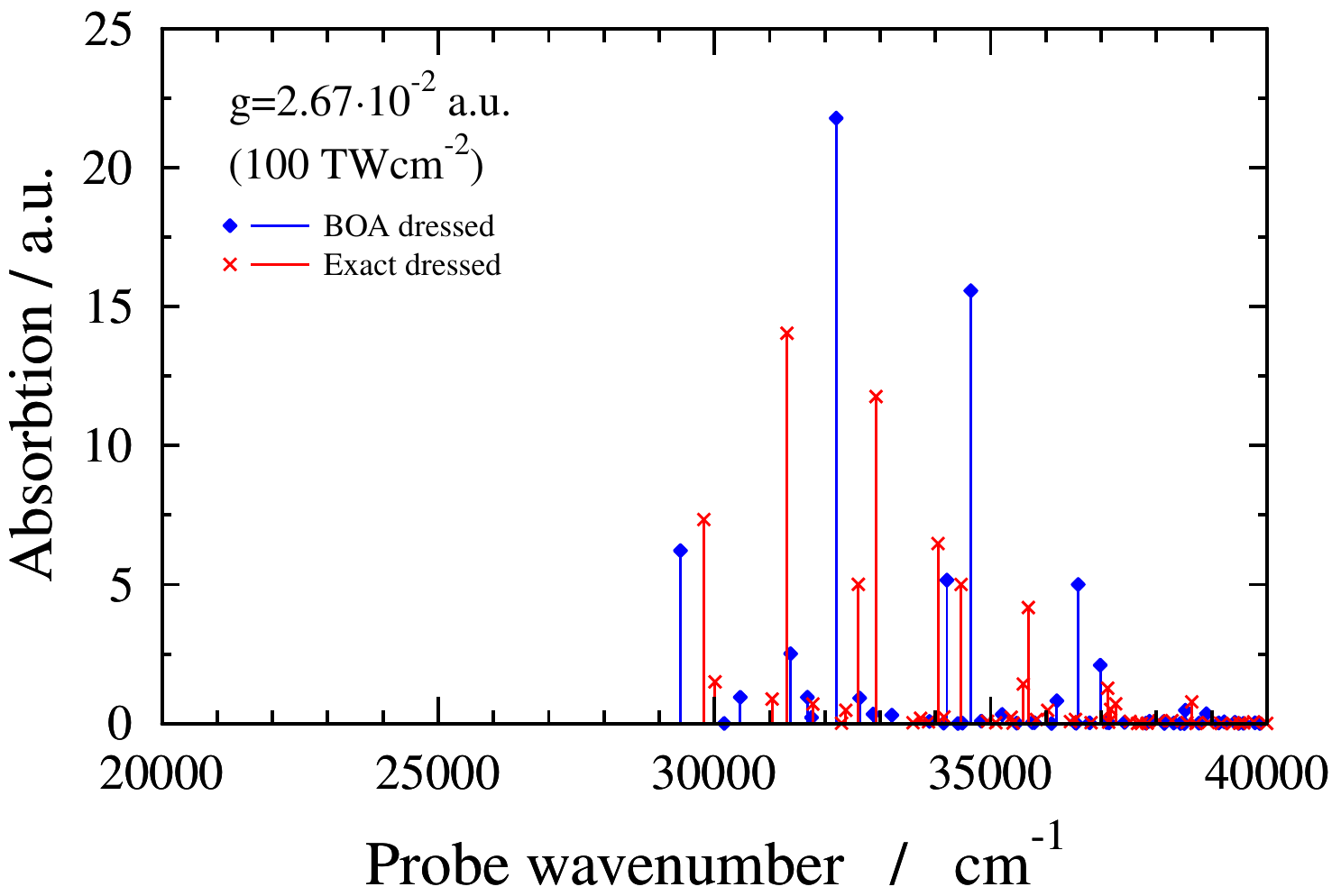}
   \includegraphics[scale=0.7]{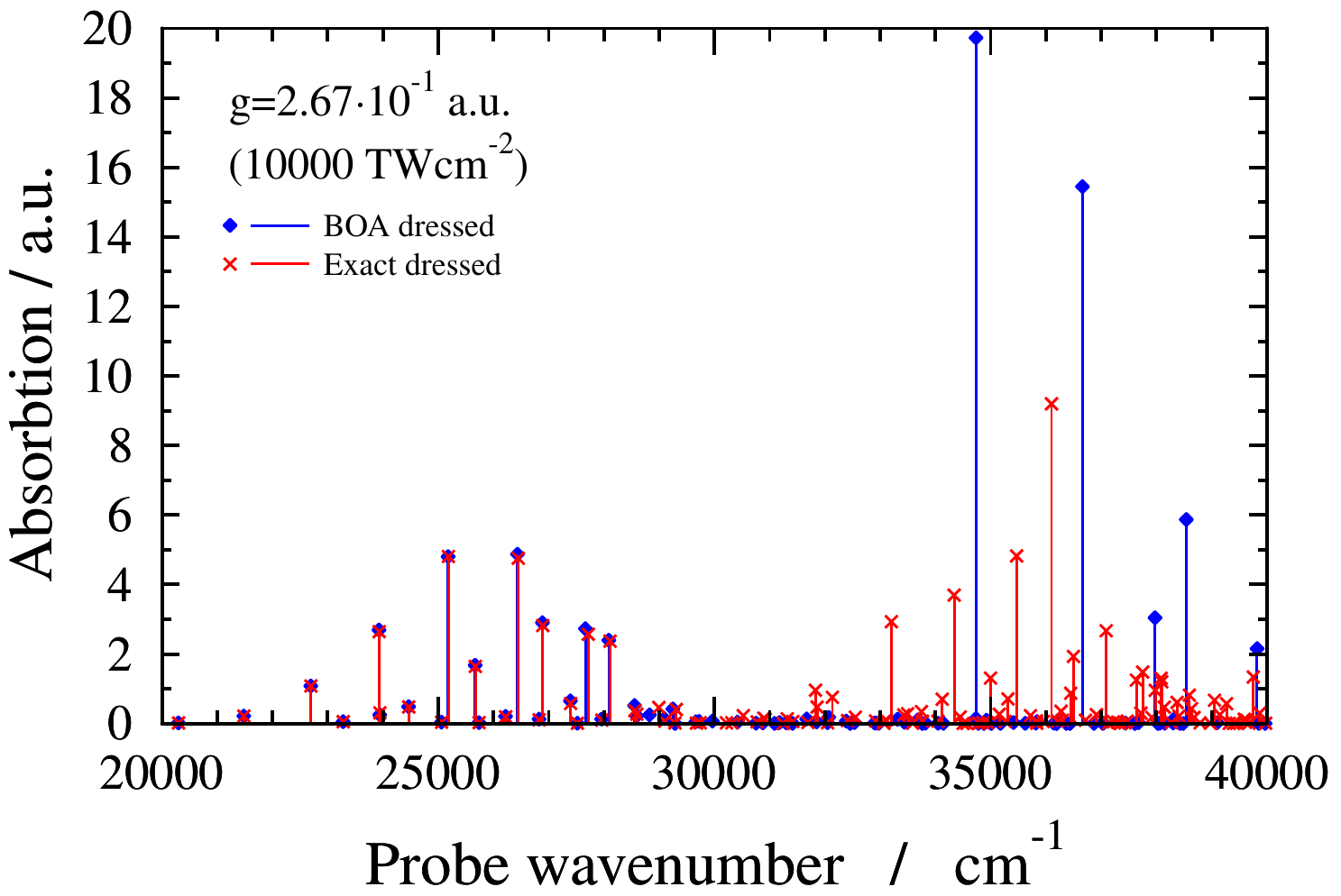}  
   \caption{2D($\nu_2,\nu_4$) dressed spectra (exact and BOA) with a cavity wavenumber of $\omega_\textrm{c} = 29957.23 ~ \textrm{cm}^{-1}$ for different coupling strength values.}
   \label{fig:spectra_2D}
\end{figure}

In this subsection results are presented for a two-dimensional quantum-dynamical model, referred to as the 2D($\nu_2,\nu_4$) model,
treating both the $\nu_2$ and $\nu_4$ vibrational modes (see the Supporting Information for technical details).
Fig. \ref{fig:spectra_2D} shows 2D($\nu_2,\nu_4$) dressed spectra ($\omega_\textrm{c} = 29957.23 ~ \textrm{cm}^{-1}$
and $\mathbf{e} = (0,1,0)$ with different $g$ values).
The exact dressed spectrum for the lowest value of $g$ in Fig. \ref{fig:spectra_2D} shows splitting of the first band of the field-free spectrum.
As $g$ increases, further field-free lines become split and shifted, and new lines appear. Common to all $g$ values is that the BOA and exact
dressed spectra differ substantially. At the highest value of $g$ one can again observe the emergence of two groups of peaks in the dressed spectrum.
In this case the BOA again works well for the lower group of peaks and utterly breaks down for the upper group of peaks.

The breakdown of the BOA in case of the 2D($\nu_2,\nu_4$) model can be attributed to the LICI, also visible in Fig. \ref{fig:molecule_cavity}.
For $\omega_\textrm{c} = 29957.23 ~ \textrm{cm}^{-1}$, the LICI is located at an energy of $30776.1 ~ \textrm{cm}^{-1}$ at $Q_2 = 8.84$ and $Q_4 = 0$.
$V_-$ and $V_+$ are degenerate at the LICI and the NAC becomes singular, also verified by analytical 2D model calculations described in the Supporting Information.
Similarly to the 1D($\nu_4$) case, the final states of the BOA transitions in the lower group of peaks for $g = 2.67 \cdot 10^{-1} ~ \textrm{au}$ are
lower-lying $V_-$ adiabatic eigenstates which have negligible amplitude at the LICI. On the other hand,
BOA lines in the upper group of peaks correspond to transitions to either higher-lying $V_-$ adiabatic eigenstates that are above the LICI in energy
or to adiabatic eigenstates of $V_+$ which has a minimum at the LICI. Therefore, the BOA can not be expected to yield sensible results for the upper group of peaks.
Note that similar conclusions have been drawn for natural CIs.\cite{81CeKoDo,Lenz1}
Along these lines, we believe that the BOA fails for the 6D model, similarly to the 2D($\nu_2,\nu_4$) model, due to the presence of the LICI.

\section{Conclusions}
This study discusses the applicability of the Born--Oppenheimer approximation (BOA) in optical cavities
for polyatomic molecules. The H$_2$CO molecule serves as a showcase example
and its absorption spectrum is calculated for the energy region of electronic transitions.
H$_2$CO does not exhibit a natural conical intersection in the studied energy
region, and therefore, nonadiabatic effects appearing in the absorption
spectrum can be attributed solely to the quantum LICI.

First, as for comparison, we have calculated the field-free spectrum of H$_2$CO utilizing
the full-dimensional (6D) as well as reduced-dimensional (2D($\nu_2,\nu_4$) and 1D($\nu_4$)) quantum-dynamical models.
It has been found that the simplest model which can approximately reproduce the structure of the
numerically-exact 6D spectrum is the 2D($\nu_2,\nu_4$) model.
As the one-dimensional 1D($\nu_2$) model leads to a vanishing absorption spectrum,
the 1D($\nu_4$) model is the one-dimensional model to be used. 

For the reduced-dimensional 1D($\nu_4$) and 2D($\nu_2,\nu_4$) models, the
field-dressed exact and BOA spectra have been computed and compared.
A striking finding of our work is that the BOA can fail even for a
one-dimensional quantum-dynamical treatment of H$_2$CO irrespective of the value of the coupling strength.
This complements previous results claiming that
the BOA can be used in the strong coupling regime when only one vibrational dof is taken into account.
Analytical considerations fully corroborate our conclusion and point out that
the breakdown of the BOA for H$_2$CO arises due to a symmetry property of the molecule.
Clearly, one should be careful even when describing a molecule in a cavity
using the BOA with only one vibration.

Moreover, we have also shown that the BOA is not applicable for the 2D($\nu_2,\nu_4$) model.
We can expect such a failure whenever there is a LICI and the spectrum extends to energies above the LICI.
Consequently, we have no doubt that the BOA fails for the full-dimensional (6D) model as well.

The above-discussed aspects are valid when the system under investigation lacks any inherent
nonadiabatic phenomena in the absence of the cavity.
If this is not the case, the situation is expected to be even more involved.

\begin{acknowledgement}
Professor Joel Bowman is gratefully acknowledged for providing Fortran subroutines for
the $\textrm{S}_0$ and $\textrm{S}_1$ potential energy surfaces.
We are indebted to Benjamin Lasorne for fruitful discussions.
This research was supported by the EU-funded Hungarian grant EFOP-3.6.2-16-2017-00005.
The authors are grateful to NKFIH for financial support (grants No. K128396 and PD124699).
\end{acknowledgement}

\bibliography{h2co_cavity_paper_arxiv}

\end{document}